\documentclass[twocolumn,showpacs,preprintnumbers,amsmath,amssymb]{revtex4}
\usepackage{graphicx}
\usepackage{dcolumn}
\usepackage{bm}
\usepackage{epsfig}

\newcommand{\be}{\begin{equation}}
\newcommand{\ee}{\end{equation}}
\newcommand{\ba}{\begin{eqnarray}}
\newcommand{\ea}{\end{eqnarray}}
\newcommand{\bi}{\begin{itemize}}
\newcommand{\ei}{\end{itemize}}
\newcommand{\tr}{{\rm Tr\,}}
\newcommand{\re}{\mathop{\rm Re}}

\newcommand{\nn}{\nonumber \\}

\newcommand{\ovl}{\overline}

\newcommand{\half}{{\textstyle\frac{1}{2}}}

\newcommand{\quarter}{{\textstyle\frac{1}{4}}}

\newcommand{\<}{\langle}
\renewcommand{\>}{\rangle}
\newcommand{\eq}{Eq.~}

\newcommand{\eqs}{Eqs.~}

\newcommand{\tab}{Tab.~}
\newcommand{\la}{\label}

\newcommand{\txts}{\textstyle}

\newcommand{\bg}{b_{\rm g}}

\begin{document}

\preprint{MIT-CTP 3840}

\title{Gluon contributions to the pion mass and light cone momentum fraction} 

\author{Harvey~B.~Meyer, John W. Negele}
\email{meyerh@mit.edu, negele@lns.mit.edu}
\affiliation{Center for Theoretical Physics\\ Massachusetts Institute of Technology\\
Cambridge, MA 02139, U.S.A.}

\date{\today}

\begin{abstract}
We calculate the matrix elements of the gluonic contributions to the 
energy-momentum tensor for a pion of mass $600 < M_\pi < 1100$ MeV in quenched lattice QCD. 
We find that gluons contribute $(37\pm 8\pm 12)\%$ of the pion's light cone momentum.
The bare matrix elements corresponding to the trace anomaly contribution to the 
pion mass are also obtained.
The discretizations of the energy-momentum tensor we use have 
other promising applications, 
ranging from calculating the origin of hadron spin to QCD thermodynamics.
\end{abstract}

\maketitle


\noindent\textit{Introduction.---~}
A striking feature of QCD is the large contribution of gluons to the mass and 
momentum of hadrons, so it is of fundamental interest to calculate the contributions of
gluons from first principles using lattice QCD.

The first moments
\ba
\<x\>_{\rm f}(q^2) &\equiv& {\txts\sum_{f=u,d,s}\int_0^1} xdx \left\{\bar f(x,q^2) + f(x,q^2)\right\} \\
\<x\>_{\rm g}(q^2) &\equiv& {\txts \int_0^1} xdx ~ g(x,q^2) 
\ea
of the quark and gluon distribution functions $f(x),~\bar f(x)$ ($f=u,d,s,\dots$) and $g(x)$ 
acquire a precise field-theoretic meaning via the 
operator product expansion in QCD. They 
satisfy the well-known momentum sum rule (MSR) $\<x\>_{\rm f}(q^2) +\<x\>_{\rm g}(q^2)=1$
and are related to the corresponding contributions to the  
energy-momentum tensor $T_{\mu\nu}$ evaluated on the hadronic state.
Separating the traceless part $\ovl T_{\mu\nu}$ from the trace part $S$ 
for gluons, denoted `g', and quarks, denoted `f',
$T_{\mu\nu}$ has the explicit form
\ba
T_{\mu\nu} \! & \equiv & \!  \overline T_{\mu\nu}^{\rm g} + 
                  \overline T_{\mu\nu}^{\rm f}
               + \quarter\delta_{\mu\nu}(S^{\rm g} + S^{\rm f}), \\
\overline T_{\mu\nu}^{\rm g}\! &=& \!
{\txts\frac{1}{4}}\delta_{\mu\nu}F_{\rho\sigma}^a F_{\rho\sigma}^a
   - F_{\mu\alpha}^a F_{\nu\alpha}^a ,\nn
\overline T_{\mu\nu}^{\rm f} \!&=& \!
\quarter {\txts\sum_f} \bar\psi_f\!\! \stackrel{\leftrightarrow}{D_{\mu}}\!\gamma_{\nu}\psi_f  
+ \bar\psi_f\!\! \stackrel{\leftrightarrow}{D_{\nu}}\!\gamma_{\mu}\psi_f  
 -{\txts\frac{1}{2}}\delta_{\mu\nu}  \bar\psi_f 
\! \stackrel{\leftrightarrow}{D_{\rho}}\! \gamma_{\rho}\psi_f ,\nn
S^{\rm g} &=& \beta(g)/(2g) ~ F_{\rho\sigma}^a  F_{\rho\sigma}^a,\quad
S^{\rm f} =  [1+\gamma_m(g)] {\txts \sum_f} \bar\psi_f m\psi_f
\nonumber
\ea
where $\stackrel{\leftrightarrow}{D_{\mu}}=
\stackrel{\rightarrow}{D_{\mu}} - \stackrel{\leftarrow}{D_{\mu}} $,
$\beta(g)$  is the beta-function, $\gamma_m(g)$  is the anomalous dimension
of the mass operator,  and all expressions are written in Euclidean space.
For an on-shell particle with four-momentum $p=(iE_p,{\bf p})$,
$E^2_{\bf p} = M^2 + {\bf p}^2$, we have the relations
\ba
 \< \Psi,{\bf p}|{\txts\int}\! d^3{\bf z}\,\overline T_{00}^{\rm f,g}(z)\, | \Psi,{\bf p}\>
\!&=&\! [E_{\bf p} -\quarter M^2/E_{\bf p}]~ \<x\>_{\rm f,g},\quad \la{eq:x}\\
 \< \Psi,{\bf p}|{\txts\int}\! d^3{\bf z}\,S^{\rm f,g}(z)\, | \Psi,{\bf p}\>
\!&=&\! (M^2/E_{\bf p}) ~b_{\rm f,g}, \la{eq:b}\\
\<x\>_{\rm f} + \<x\>_{\rm g} &=& b_{\rm f} + b_{\rm g} = 1,  \la{eq:xb}
\ea
where states are normalized according to $\<{\bf p}|{\bf p} \> = 1 $. 
We shall return to the renormalization of $\<x\>_{\rm f,g}$ below.

Equation~\ref{eq:x}  shows that in the infinite momentum frame, where $E_p \sim P \to \infty$, 
$\langle x \rangle_g$ represents the momentum fraction arising from  gluons, 
and calculating $\langle x \rangle_g$ is the main goal of this work.  
In the rest frame, the gluon contribution of Eq.~\ref{eq:x} to the hadron mass 
is $\frac{3}{4} M   \langle x \rangle_g$~\cite{ji}.   From Eq.~\ref{eq:b} in the rest frame,
the contribution of the trace anomaly $S^g$ to the hadron mass is 
$\frac{1}{4}b_g M$~\cite{ji}, and in this work 
we perform the first step to calculate this matrix element as well.

Whereas  non-singlet matrix elements can now be calculated to high precision in full QCD in the
chiral regime~\cite{Edwards:2005ym,Gockeler:2006ns,Hagler:2007xi}, 
calculations of matrix elements of singlet operators are far less developed due 
to the computational challenges of calculating 
disconnected diagrams, which require all-to-all propagators, 
and matrix elements of gluon fields, which are notoriously noisy due to quantum fluctuations.
The first attempt to calculate the quark momentum fraction was 
in the proton in~\cite{horsley}, and was  found to be numerically very challenging.
In this exploratory study we treat the case of ``heavy pions'' with masses in the range
$600\, {\rm MeV}<M_\pi<1060 \,{\rm MeV}$, where hadronic matrix elements 
in the quenched approximation,  which neglects quark loops, are generally 
close to those in  full QCD. 
The techniques developed here are applicable 
in full QCD calculations, and to the case of the proton.

\noindent\textit{Lattice formulation.---~}
 We use the Wilson gluon action
$ \frac{1}{g_0^2} \sum_{x,\mu\neq\nu} \tr\{1-P_{\mu\nu}(x)\}$, 
where $P_{\mu\nu}$ is the plaquette, and the Wilson fermion action~\cite{wilson74} at an
inverse coupling $ 6/g_0^2 \equiv \beta=6.0$, corresponding to a lattice spacing
$a=0.093\,$fm  for $r_0=0.5\,$fm~\cite{necco-sommer}.
There are two distinct ways~\cite{liu} to discretize the Euclidean gluon energy operator
$\overline T^{\rm g}_{00} = \half (-{\bf E}^a \cdot {\bf E}^a + {\bf B}^a \cdot {\bf B}^a)$
and the trace anomaly 
\mbox{$S^{\rm g} = \frac{\beta(g)}{g} ({\bf E}^a \cdot {\bf E}^a + {\bf B}^a \cdot {\bf B}^a)$ }
on a hypercubic lattice.

The first, denoted `bp' for bare-plaquette,  uses a sum of bare plaquettes 
$P_{\mu\nu}$ around a body-centered point $x_\odot = x + \half a \sum_\mu \hat\mu$,
 which, when summed over a time slice, yields 
\ba
&& a^3\sum_{\bf x}
 \overline T_{00}^{\rm bp}(x_\odot) 
= \frac{2\chi^{\rm bp}(g_0)Z_{\rm g}(g_0)}{ag_0^2}\sum_{\bf x}\la{eq:T00plaq}  \\
&& \re\tr \Big[ {\txts\sum_{k}}  P_{0k}(x) - {\txts\sum_{k<l}}~ \half[P_{kl}(x)+P_{kl}(x+a\hat0)]  \Big],
\nonumber \\
&& a^3\sum_{\bf x}
 S^{\rm bp}(x_\odot) 
= \frac{2 \chi_s^{\rm bp}(g_0)}{a} \frac{dg_0^{-2}}{d\log a} \sum_{\bf x} \re\tr \times \nn
&& \Big[ {\txts\sum_{k}} ( 1 - P_{0k}(x) ) +  {\txts\sum_{k<l}} (1 -\half[P_{kl}(x)+P_{kl}(x+a\hat0) ]) \Big].
\nonumber
\ea

The other form, denoted `bc' for bare clover, is 
\ba
\overline T_{00}^{\rm bc}(x)\!\!\!
&\equiv& \!\!\!\frac{\chi^{\rm bc}(g_0) Z_{\rm g}(g_0)}{g_0^2}
\re\!\tr \!\Big[{\txts\sum_{k}} (\widehat F_{0k})^2 
     \! -  \!  {\txts\sum_{k<l}} (\widehat F_{kl})^2   \Big]  
\la{eq:T00clover} \\
S^{\rm bc}(x) \!\! & \equiv &\!\! \chi^{\rm bc}_s(g_0) \frac{dg_0^{-2}}{d\log a}
\re\!\tr\!\Big[{\txts\sum_{k}} (\widehat F_{0k})^2
      +  {\txts\sum_{k<l}}(\widehat F_{kl})^2   \Big],\,\,
\nonumber
\ea
where $\widehat F_{\mu\nu}(x)$ is the clover-shaped discretization of 
the field-strength tensor (see~\cite{sommer96}). 
This form allows for the discretizations
of off-diagonal elements of $\overline T_{\mu\nu}$ as well.
Each of the normalization factors $Z_{\rm g}(g_0)$, $\chi^{\rm bc}(g_0)$ 
and $\chi^{\rm bc}_s(g_0)$ 
in \eq (\ref{eq:T00plaq},\ref{eq:T00clover}) is of the form
$1+{\rm O}(g_0^2)$.

An additional freedom in discretization is local smoothing of the fields
by replacing each link in \eqs(\ref{eq:T00plaq},\ref{eq:T00clover}) by a 
sum of a connected product of links joining the same two lattice points.
This only changes the fields by higher dimension operators, and HYP 
smearing~\cite{hyp} is particularly suited for this application because it
preserves the symmetry between all Euclidean directions and is localized within 
a single hypercube. We use the original HYP-smearing parameters~\cite{hyp}, and  project
onto SU(3) as in~\cite{hqet-actions}.

Our criteria for the choice of the discretization are 
to maximize the signal-to-noise ratio, minimize cutoff effects,  
and preserve locality as much as possible.
The noisiest quantity we calculate is $\ovl T_{00}(x)$, which involves the near cancellation of
${\bf E}^2$ and ${\bf B}^2$.  Hence, we studied the signal-to-noise ratio for four different discretizations
by comparing the variance of a related thermodynamic variable, the entropy density at 
temperature $T=1/L_0=1.21T_c$~\cite{teper-sun}, which is proportional to the 
expectation value of   $\sum_x \ovl T_{00}(x)$, on an $L_0\times L^3 $ lattice with 
$L/a=16$ and $L_0/a =6$. The resulting variances  for the plaquette and clover discretizations
with bare and HYP links are shown in Table \ref{tab:var}.
We find  dramatic differences between the discretizations, with
HYP smearing reducing the bare plaquette variance by a factor of 41 
and the HYP-clover operator reducing the variance by a factor of 87.
Variance reduction comes at the cost of a certain loss of locality, 
since the HYP plaquette and HYP-clover operators have extent $3a$ and $4a$ respectively.

\begin{table}
\begin{center}
\begin{tabular}{|c|c|c@{~~~}c|c@{~~~}c|}
\cline{3-6}
\multicolumn{2}{c|}{}& \multicolumn{2}{c |}{relative variance}  &  \multicolumn{2}{c|}{normalization} \\
\cline{3-6}
\multicolumn{2}{c|}{}    & bare  &  HYP & bare  &  HYP \\
\hline
$\ovl T_{00}$& plaq.   & 26.4(71)    & 0.6518(43) &   1         & 0.5489(68) \\
& clover  & 3.85(11)    & 0.3049(41) &  2.184(67)  & 0.613(20)  \\  
\hline
$S$& plaq.   & 2.64 (12)  &  0.474(13) &  1        &  0.9951(77)   \\
& clover  & 1.180(39)  &  0.2975(72) &   4.062(30)   &   1.410(13)   \\
\hline
\end{tabular}
\end{center}
\caption{\underline{Left}: the relative variance, $\<{\cal O}^2\>/\<{\cal O}\>^2-1$,
of the operators ${\cal O}=\sum_x ( o(x)-\< o\>_0)$ (top: $o=\ovl T_{00}$,
bottom:  $o=S$) on a $6\times16^3$ lattice at $\beta=6.0$
for different discretizations  described in the text. 
\underline{Right}:
the normalization $\chi(g_0,a/L_0)$ (top) and $\chi_s(g_0,a/L_0)$  (bottom)
of the operator relative to the bare plaquette, determined on the same lattice.}
\la{tab:var}
\end{table}

The normalization factor $Z_{\rm g}(g_0)$ appearing in \eq\ref{eq:T00plaq}
is dictated by an exact lattice sum-rule for the Wilson gauge action and 
is known with a precision of about $1\%$ (see~\cite{hm-visco} and Refs. therein).
To obtain the absolute normalization of other discretizations, it is sufficient
to compute their normalization $ \chi(g_0)$ 
relative to that of the bare plaquette, and the resulting $\chi$'s 
 are given in \tab\ref{tab:var} for the four discretizations.

As a compromise  between locality and variance reduction, from now on 
we work with the HYP-plaquette discretization.
We performed a check of its discretization errors by computing 
the dependence of $\chi$ on $a/L_0$, which is a nonlocality effect.
Figure~\ref{fig:hyp} shows that the dependence 
of $\chi$ on $a/L_0$ is mild and statistically consistent with zero
for $L_0/a\geq6$, and that
all four lattice operators are viable discretizations of the same continuum operator.
As a  check of the correct normalization of the chosen HYP-plaquette
operator, we computed its expectation value on the lightest scalar glueball. 
In that case, we know that the momentum fraction carried by the glue is one
(see~\cite{michael-tickle} for an early calculation in SU(2) gauge theory),
and indeed we find $\<x\>^{(G)}_{\rm g} =1.16(18)$.

\begin{figure}
\begin{center}
\psfig{file=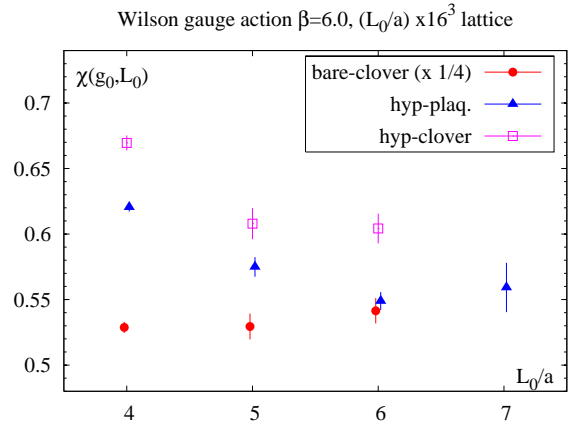,angle=-90,width=8.5cm}
\end{center}
\vspace{-0.5cm}
\caption{A study of cutoff effects: the normalization 
$\chi(g_0,a/L_0)$ of three discretizations of $\ovl T_{00}$
relative to the one based on the bare plaquette as a function of $L_0/a$.}
\la{fig:hyp}
\end{figure}

\noindent\textit{The gluon momentum fraction in the pion.---~}
We consider a triplet of Wilson quarks, labeled $u,d,s$, with periodic 
boundary conditions in all directions and with common $\kappa=0.1515,~0.1530$
and $0.1550$ corresponding to pion masses approximately 1060,
890  and 620 MeV on lattices $32\cdot12^3$, $32\cdot16^3$, $48\cdot16^3$  and $24^4$.
To calculate the gluonic momentum fraction in the pion,
we define the effective momentum fraction
\ba\la{eq:xeff}
\<x\>^{(\pi)}_{\rm g,eff}(x^{\rm min}_0) \equiv 
\frac{8}{3M_\pi}~ \frac{a^3}{|\Lambda_0|}~ \times  \hspace{3.4cm} \\
\sum_{{\bf x};\,x_0\in\Lambda_0}
\left[\frac{   \sum_{\bf y} \<j(0)~ \overline T^{\rm hp}_{00}(x_\odot)~j(\frac{L_0}{2},{\bf y})\>}
     {\sum_{\bf y'} \<j(0)~j(\frac{L_0}{2},{\bf y}')\> }
- \<\overline T^{\rm hp}_{00}(x_\odot) \>
\right],
\nonumber
\ea
and similarly for $b^{({\rm bare})}_{\rm g}$ by substituting $\overline T^{\rm hp}_{00}\to S^{\rm hp}$.
Here $\Lambda_0 \!= \!\{x_0^{\rm min},\dots,\frac{L_0}{2} \! -\! x_0^{\rm min}\! -\! a,
 \frac{L_0}{2}+x_0^{\rm min},\dots,L_0-x_0^{\rm min}\! -\! a\}$.
This corresponds to creating a pion at the origin, annihilating it at the middle time slice, 
measuring the gluon operator over all times at least $x_0^{\rm min}$ away from the source or sink, dividing by
the corresponding pion two-point function, and subtracting the vacuum expectation value of the operator.
For large $L_0$ and $ x_0^{\rm min}$, $\<x\>^{(\pi)}_{\rm g,eff} \to \<x\>^{(\pi)}_{\rm g}$.

As a source field for the pion, we use the isovector pseudoscalar density
$j(x)=\bar d(x)\gamma_5u(x)$. Its two-point function is positive 
on every configuration, for each of which we do 12 inversions 
corresponding to Dirac and color indices.
On a $24^4$ lattice, we take advantage of the symmetry between all directions
to perform these inversions at the points $k(6,6,6,6)$ for $k=0,1,2,3$
and symmetrize expression (\ref{eq:xeff}) with respect to all directions, so that
$\sum_{x,\mu}\overline T_{\mu\mu}(x)$ vanishes on every  configuration. Figure~\ref{fig:3ptfct} shows
our stable plateaus for $\<x\>^{(\pi)}_{\rm g,eff}$ at large values of $ x_0^{\rm min}$ for
two lattice sizes, and all 
the results are summarized in \tab\ref{tab:xg}.
\begin{figure}
\begin{center}
\psfig{file=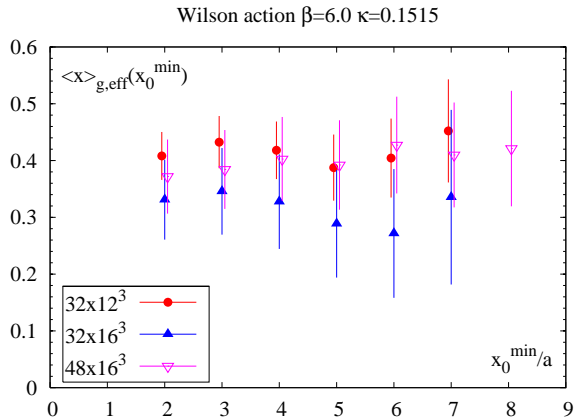,angle=-90,width=8.5cm}
\end{center}
\vspace{-0.5cm}
\caption{The effective gluonic momentum fraction, \eq\ref{eq:xeff},  in a heavy pion, 
$M_\pi \simeq 1060$MeV. }
\la{fig:3ptfct}
\end{figure}

\begin{table}
\begin{center}
\begin{tabular}{c|c@{~~~}c@{~~~}c@{~~~}c}
  $M_\pi$ (Mev)      &  $32\cdot12^3$  & $32\cdot16^3$& $48\cdot16^3$ & $24^4$  \\
\hline
1060(10)   & $0.39(6)_{23091}$   & $0.29(9)_{7113}$ & $0.40(8)_{8331}$  & $0.34(9)_{1048}$ \\
891(9)   &  ---      &  ---   &  ---      & $0.36(8)_{3066}$  \\  
624(6)  & ---       &  ---   &  ---      & $0.58(16)_{2538}$   \\
\hline
1060(10)    & $0.89(3)_{23091}$   & $0.95(5)_{7113}$ &  $1.00(4)_{8331}$ &  $0.77(7)_{1048}$    \\
891(9)  &  ---      &  ---   &  ---              &  $1.02(6)_{3066}$ \\
624(6)   & ---       &  ---   &  ---              & $ 2.2(1)_{2538}$ \\
\end{tabular}
\end{center}
\caption{The glue momentum fraction $\<x\>_{\rm g}^{(\pi)}$ (top)
and the bare trace anomaly matrix element 
$b^{\rm (bare)}_{\rm g}$ (bottom) in the pion. The integer in each subscript
denotes the number of configurations used.}
\la{tab:xg}
\end{table}

Equation \ref{eq:xb} 
has been derived for QCD at finite lattice spacing in~\cite{hm-sumrules}.
In particular we have
\be
1 = \<x\>_{\rm g} +  \<x\>_{\rm f},
\qquad  \<x\>_{\rm f} = Z_{\rm f}(g_0) \<x\>^{\rm bare}_{\rm f},
\la{eq:bare sr}
\ee
where, disregarding  disconnected diagrams,
$\<x\>^{\rm bare}_{\rm f}$ has been computed in~\cite{pion matrix element} at the 
same bare parameters $(\beta=6,\kappa=0.1530)$. The factor $Z_{\rm f}(g_0)$ 
is the fermion analog of $Z_{\rm g}(g_0)$, see \eq(\ref{eq:T00plaq}).

\noindent\textit{Renormalization of $\<x\>_{\rm g}$.---~}
Recall that, in QCD, 
the renormalization pattern in the singlet sector reads~\cite{ji-prd}
\be
\left[\begin{array}{c} \ovl T_{00}^{\rm g}(\mu) \\ \ovl T_{00}^{\rm f}(\mu)  \end{array}\right]
= \Bigg[\begin{array}{l@{~~}r} Z_{\rm gg} & 1 \!-\! Z_{\rm ff} \\
                                1\! -\! Z_{\rm gg} & Z_{\rm ff} \end{array}\Bigg]
\left[\begin{array}{c} \ovl T_{00}^{\rm g}(g_0) \\ \ovl T_{00}^{\rm f}(g_0) \end{array}\right],
\ee
provided $\ovl T_{00}^{\rm f,g}(g_0)$ are normalized so that
Eqs.~(\ref{eq:x},\ref{eq:xb}) hold.
In lattice regularization, this requires the scheme-independent
$Z_{\rm g}(g_0)$ and $Z_{\rm f}(g_0)$ factors, while $Z_{\rm gg}$
and $Z_{\rm ff}$ are scheme-dependent functions of $(a\mu,g_0)$.
The renormalization group equation then takes the form
\be
\mu\partial_\mu \!
\left[\begin{array}{c}\! \<x\>_{\rm g}(\mu^2)\! \\ \! \<x\>_{\rm f}(\mu^2)\!  \end{array}\right]
\! =-\bar g^2(\mu)\! \left[\begin{array}{l@{~~}r} c_{\rm gg}(\bar g) & - c_{\rm ff}(\bar g) \\
- c_{\rm gg}(\bar g) & c_{\rm ff}(\bar g) \end{array}\right] \!\!
\left[\begin{array}{c}\! \<x\>_{\rm g}(\mu^2)\! \\ \! \<x\>_{\rm f}(\mu^2)\!  \end{array}\right]
\nonumber
\ee
with $ \mu\partial_\mu \log[Z_{\rm gg} + Z_{\rm ff}-1] = -\bar g^2[ c_{\rm gg}+ c_{\rm ff}]$
and $c_{\rm gg,ff}(\bar g=0)= \frac{N_{\rm f}}{12\pi^2}, ~ \frac{4}{9\pi^2}$ 
respectively~\cite{gross-wilczek,georgi-politzer}.
Besides the zero-mode $\ovl T_{00}$, the linear combination
$[1+\tau(\mu)]\ovl T^{\rm g}_{00}(\mu)+\tau(\mu) \ovl T^{\rm f}_{00}(\mu)$
renormalizes multiplicatively with anomalous dimension $-\bar g^2[c_{\rm ff}+ c_{\rm gg}]$, 
where $\mu\partial_\mu\tau = -\bar g^2[(c_{\rm ff}+ c_{\rm gg}) \tau + c_{\rm ff}]$.
Note that the asymptotic glue momentum fraction
is given by $c_{\rm ff}(0)/[c_{\rm ff}(0)+c_{\rm gg}(0)]
= Z_{\rm gg}(\infty)=1\! -\! Z_{\rm ff}(\infty)= -\tau(\infty)=16/[16+3N_{\rm f}]$.

In the quenched approximation, $Z_{\rm gg}=1$ due to the absence
of quark loops~\cite{singlet MSbar,singlet SF}. This implies that the singlet part renormalizes 
multiplicatively and with the same anomalous dimension as
the non-singlet part, which has been computed 
non-perturbatively in~\cite{continuous external momenta},
\ba
\<x\>_{\rm g}(\mu^2) &=& \<x\>_{\rm g} + [1-Z_{\rm ff}(a\mu,g_0)] ~\<x\>_{\rm f},\\
\<x\>_{\rm f}(\mu^2) &=& Z_{\rm ff}(a\mu,g_0) ~ \<x\>_{\rm f}.
\ea
The factor $Z_{\rm f}(g_0)=1+{\rm O}(g_0^2)$
is as yet unknown beyond tree level. If we allow for a conservative error, 
based on the typical size of one-loop corrections, $Z_{\rm f}(g_0)=1.0(2)$,
then using $\<x\>^{\rm bare}_{\rm f}= 0.616(4)$ and
$Z_{\rm ff}(a\mu,g_0)Z_{\rm f}(g_0)=0.99(4)$ for the $\ovl{MS}$-scheme 
at $\mu=2$GeV~\cite{continuous external momenta,pion matrix element}, 
our final result is
\be
\<x\>^{(\pi)}_{\rm g}(\mu^2_{\ovl{MS}}=4{\rm GeV}^2) = 0.37(8)(12) \qquad (M_\pi=890{\rm MeV}),
\nonumber
\ee
where the first error is statistical and the second comes from the uncertainty in
$Z_{\rm f}(g_0)$.

Finally, our result for the glue momentum fraction in a (heavy) pion 
is compatible with
phenomenological determinations~\cite{SMRS92,gluck}, 
 $\<x\>^{\ovl{MS}}_g=0.38(5)$ at $Q^2=4{\rm GeV}^2$, based on Drell-Yan,
prompt photoproduction, and the model assumption 
that sea quarks carry 10-20\% of the momentum. 
The agreement suggests a mild quark-mass dependence, but only a calculation in
full QCD and at smaller masses can substantiate this. Our result at $Q^2=4{\rm GeV}^2$ lies
clearly below the $N_{\rm f}=3$ asymptotic glue momentum fraction of 0.64.
The fact that our result and the valence quark momentum fraction,
computed in~\cite{pion matrix element}, add up to $0.99(8)(12)$ 
suggests that the omitted disconnected diagrams are small.

\noindent\textit{Discussion of $b_{\rm g}$.---~}
In a chirally symmetric formulation of massless QCD, the trace anomaly is the only contribution 
to $S(x)$, and its matrix elements are renormalization group invariant.
With Wilson fermions however, the absence of chiral symmetry implies that 
the trace anomaly acquires a 
linearly divergent contribution from the operator $\bar \psi \psi$. Thus our matrix 
elements $\bg^{\rm (bare)}$ should be regarded as  intermediate results.
The coefficient of the counterterm, as well as its disconnected diagrams,
will have to be computed before we can quote a physical value for $b_{\rm g}$ in the pion.
Not surprisingly,  $b^{\rm (bare)}_{\rm g}$ shows a strong quark-mass dependence,
since the missing disconnected diagrams are suppressed by $1/m$.
We note that $b^{\rm (bare)}_{\rm g} \sim 0.9 (1) $ at the largest mass 
is of the same order of magnitude as Ji's phenomenological estimate 
of $b_{\rm g}$ in the proton~\cite{ji}, $0.85(5)$.

\noindent\textit{Conclusion.---~}
We have computed the glue momentum fraction $\<x\>_{\rm g}$ 
in a pion of mass $0.6{\rm GeV}<M_\pi<1.06{\rm GeV}$ using quenched lattice QCD
simulations. We find $37(8)(12)\%$ at $\mu_{\ovl{MS}}=2$GeV,
a result compatible with phenomenological determinations~\cite{SMRS92,gluck}.

Although it appears difficult to achieve 
precision at the percent level, the present method is applicable to full QCD with dynamical quarks.
Presently the larger uncertainty comes from the 
normalization of the quark contribution to the renormalized $\<x\>_{\rm g}$,
and could be reduced significantly by a one-loop calculation.

We also evaluated the bare trace anomaly contribution to the pion's mass
in the same framework. 
The counterterm remains to be calculated, but it will ultimately be preferable 
to use chiral fermions to avoid mixing with the lower dimensional fermion operator.

Finally, we remark that 
the freedom of choosing a numerically advantageous discretization 
of $\ovl T_{\mu\nu}$ has not been fully exploited 
in previous lattice simulations.
The improvement that was essential in the present computation of the pion
momentum fraction can be carried over 
to fully dynamical calculations and the exploration of other observables,
such as the gluon contribution to the nucleon spin.   It is also particularly
promising for thermodynamic studies of pressure, 
energy density and transport coefficients.

\noindent\textit{Acknowledgments.---~~}
We thank Bob Jaffe and Frank Wilczek for stimulating discussions,
and Andrea Shindler and Stefano Capitani for a correspondence
on existing literature. This work was supported in part by 
funds provided by the U.S. Department of Energy under cooperative research agreement
DE-FG02-94ER40818.



\begin{thebibliography}{99}


\bibitem{ji}
  X.~D.~Ji,
  Phys.\ Rev.\ Lett.\  {\bf 74}, 1071 (1995)
  [arXiv:hep-ph/9410274].
  
\bibitem{Edwards:2005ym}
  R.~G.~Edwards {\it et al.}  [LHPC Collaboration],
  Phys.\ Rev.\ Lett.\  {\bf 96}, 052001 (2006)
  [arXiv:hep-lat/0510062].

\bibitem{Gockeler:2006ns}
  M.~Gockeler {\it et al.},
  PoS {\bf LAT2006}, 179 (2006)
  [arXiv:hep-lat/0610066].

\bibitem{Hagler:2007xi}
  Ph.~Hagler {\it et al.}  [the LHPC and MILC Collaborations],
  arXiv:0705.4295 [hep-lat].

  

\bibitem{horsley}
  M.~Gockeler {\it et al.},
  Nucl.\ Phys.\ Proc.\ Suppl.\  {\bf 53}, 324 (1997)
  [arXiv:hep-lat/9608017].

\bibitem{wilson74}
  K.G.~Wilson,
  Phys.\ Rev.\  D {\bf 10}, 2445 (1974).

\bibitem{necco-sommer}
  S.~Necco and R.~Sommer,
  Nucl.\ Phys.\ B {\bf 622} (2002) 328.

\bibitem{liu}
  Y.~Chen {\it et al.},
  Phys.\ Rev.\  D {\bf 73}, 014516 (2006)
  [arXiv:hep-lat/0510074].

\bibitem{teper-sun}
  B.~Lucini, M.~Teper and U.~Wenger,
  JHEP {\bf 0401}, 061 (2004).

\bibitem{sommer96}
  M.~Luscher, S.~Sint, R.~Sommer and P.~Weisz,
  Nucl.\ Phys.\  B {\bf 478}, 365 (1996)
  [arXiv:hep-lat/9605038].

\bibitem{hyp}
  A.~Hasenfratz and F.~Knechtli,
  Phys.\ Rev.\  D {\bf 64}, 034504 (2001)
  [arXiv:hep-lat/0103029].

\bibitem{hqet-actions}
  M.~Della Morte, A.~Shindler and R.~Sommer,
  JHEP {\bf 0508}, 051 (2005)
  [arXiv:hep-lat/0506008].

\bibitem{hm-visco}
  H.~B.~Meyer,
  arXiv:0704.1801 [hep-lat].

\bibitem{michael-tickle}
  G.~A.~Tickle and C.~Michael,
  Nucl.\ Phys.\  B {\bf 333}, 593 (1990).

\bibitem{hm-sumrules}
  H.~B.~Meyer,
  Nucl.\ Phys.\  B {\bf 760}, 104 (2007)
  [arXiv:hep-lat/0609007].

\bibitem{ji-prd}
  X.~D.~Ji,
  Phys.\ Rev.\  D {\bf 52}, 271 (1995)
  [arXiv:hep-ph/9502213].

\bibitem{gross-wilczek}
  D.~J.~Gross and F.~Wilczek,
  Phys.\ Rev.\  D {\bf 9}, 980 (1974).

\bibitem{georgi-politzer}
  H.~Georgi and H.~D.~Politzer,
  Phys.\ Rev.\  D {\bf 9}, 416 (1974).

\bibitem{continuous external momenta}
  M.~Guagnelli, K.~Jansen, F.~Palombi, R.~Petronzio, A.~Shindler and I.~Wetzorke
                  [Zeuthen-Rome / ZeRo Collaboration],
  Nucl.\ Phys.\  B {\bf 664}, 276 (2003)
  [arXiv:hep-lat/0303012].

\bibitem{pion matrix element}
  M.~Guagnelli, K.~Jansen, F.~Palombi, R.~Petronzio, A.~Shindler and I.~Wetzorke
                  [Zeuthen-Rome (ZeRo) Collaboration],
  Eur.\ Phys.\ J.\  C {\bf 40}, 69 (2005)
  [arXiv:hep-lat/0405027].

\bibitem{singlet SF}
  F.~Palombi, R.~Petronzio and A.~Shindler,
  Nucl.\ Phys.\  B {\bf 637}, 243 (2002)
  [arXiv:hep-lat/0203002].

\bibitem{singlet MSbar}
  S.~Capitani and G.~Rossi,
  Nucl.\ Phys.\  B {\bf 433}, 351 (1995)
  [arXiv:hep-lat/9401014].

\bibitem{SMRS92}
  P.~J.~Sutton, A.~D.~Martin, R.~G.~Roberts and W.~J.~Stirling,
  Phys.\ Rev.\  D {\bf 45}, 2349 (1992).


\bibitem{gluck}
  M.~Gluck, E.~Reya and I.~Schienbein,
  Eur.\ Phys.\ J.\  C {\bf 10}, 313 (1999)
  [arXiv:hep-ph/9903288].
\end{thebibliography}
\end{document}